\documentclass[useAMS,usenatbib]{mn2e}
\usepackage[dvips]{graphicx}
\usepackage{epsfig}
\usepackage{longtable}
\usepackage{setspace}
\usepackage{amssymb}
\newcommand{\suzaku}{{\it Suzaku }}

\begin{document}

\title{Broadband Spectral Analysis of Aql X-1}

\author[Raichur, Misra \& Dewangan]{Harsha Raichur\thanks{E-mail: harsha@iucaa.ernet.in}, Ranjeev Misra  and Gulab Dewangan \\
Inter-University Centre for Astronomy and Astrophysics,  Post Bag 4, Ganeshkhind, Pune-411007, India }

\date{Accepted.....; Received .....}
\maketitle

\begin{abstract}

We present the results of a broadband spectral study of the
transient Low Mass X-ray Binary Aql X-1 observed by \suzaku and
{\it Rossi X-ray Timing Explorer} satellites. The source was observed
during its 2007 outburst in the High/Soft (Banana) state and in the
Low/Hard (Extreme Island) state. Both the Banana state  and the
Extreme Island state spectra are best described by a two component
model consisting of a soft multi-colour blackbody emission likely 
originating from the accretion disk and a harder Comptonized emission 
from the boundary layer. Evidence for a hard tail (extending to $\sim 50$ 
keV) is found during the Banana state; this further (transient) component, 
accounting for atleast $\sim 1.5$\% of the source luminosity, is modeled by a 
power-law. Aql X-1 is the second Atoll source after GX 13+1 to show 
a high energy tail. The presence of a weak but broad Fe line provides 
further support for a standard accretion disk extending nearly to the neutron star
surface. The input photons for the Comptonizing boundary layer could
either be the disk photons or the hidden surface of the star or
both. The luminosity of the boundary layer is similar to the disk
luminosity in the banana state and is about six times larger in the
extreme island state. The temperature of the Comptonizing boundary
layer changes from $\sim 2$ keV in the banana state to $\sim 20$ keV
in the extreme island state.

\end{abstract}


\section{Introduction}

Low mass X-ray binaries (LMXB) containing weakly magnetised neutron
stars (NS) are classified into Z and Atoll sources \citep{Has89,Van06}. 
The classification is based upon the different tracks these two types of 
sources trace in the colour-colour diagrams (CD) and their associated 
timing properties in the different spectral states.
Even though detailed timing and spectral differences between the two
class of sources are not completely explained, higher accretion rate is 
generally believed to characterise the Z as compared to the Atoll sources.
Atoll sources have lower luminosities (~0.001-0.5 $L_{Edd}$) than 
Z sources and they have fragmented CD consisting of banana
state (BS) and Extreme island state (EIS). There is also
a transitional state between these two states referred to as 
Island state. In the BS the source luminosity is high
and the observed energy spectrum is soft (high/soft state). 
On the other hand in the EIS the source luminosity is low
and the observed energy spectrum is hard (low/hard state).

Irrespective of the type of source, there are two different 
approaches to model their energy spectra, both consisting of a 
soft/thermal and a hard/Comptonized component 
\citep[see][for a review]{Bar01}. In one approach the accretion disk is
considered to be emitting the soft thermal component while the boundary layer
emits the hard Comptonised component \citep[``Eastern Model''][]{Mit89}. In the
other approach, the boundary layer emits the soft thermal component
and the inner hot accretion disk emits the hard Comptonised component
of the spectrum \citep[``Western Model''][]{Whi88}. To distinguish between these two 
approaches good quality broadband spectral data is
required. Beppo-SAX broadband (0.1-200 keV) observations show that
Eastern-like models are preferred for several NS-LMXB 
\citep[e.g.][]{Gua98,Lav04,Tar08}. 

\emph{Suzaku} observations cover a broad energy range of
0.7-70 keV with good spectral resolution. This capability of \suzaku allows to
distinguish between several modelling approaches for NS-LMXBs: Eastern-like, as
in the case of the Z-source LMC X-2 \citep{Agr09}; Western-like, as in the case
of the periodic burster XB1323-619 \citep{Bal09}; scattered-in dust emission 
plus direct NS and/or disk blackbody emission, as for the eclipsing LMXB SAXJ1745.6-2901
\citep{Hyo09}.

\begin{table*}
\begin{minipage}{180mm}
\begin{center}
\caption{Suzaku observation details}
~\\
\begin{tabular}{lllllll}
\hline
Suzaku& Suzaku Obs & Suzaku Obs & \multicolumn{2}{c}{Source count rate}& Corresponding RXTE & RXTE Obs \\
Obs Id&Start time (MJD) & Exposure (s)&\multicolumn{2}{c}{from Suzaku obs(cnt/s)$^{\dag}$} & Obs Id & Start time (MJD)\\
      &                 &             & XIS & PIN & & \\
\hline
402053010 (BS) & 54371.64 & 13825 & $139.90 \pm 0.18$ & $1.13 \pm 0.01$ & None           & - \\
402053020 (EI1)& 54376.99 & 15132 & $13.27 \pm 0.02$  & $0.79 \pm 0.01$ & 93405-01-04-01 & 54376.98\\
402053030 (EI2)& 54382.21 & 19711 & $15.77 \pm 0.02$  & $0.96 \pm 0.01$ & 93405-01-05-01 & 54382.34\\
402053040 (EI3)& 54388.33 & 17915 & $12.10 \pm 0.02$  & $0.79 \pm 0.01$ & 93405-01-06-00 & 54388.36\\
\hline
\end{tabular}
\end{center}
$^{\dag}$ Errors quoted on count rates are 1 sigma errors
\label{tab:table1}
\end{minipage}
\end{table*}

Aql X-1 is a recurrent soft transient LMXB with an optical companion star of V =
21.6 K7V spectral type \citep{Che99}  and orbital period 18.9
hours \citep{Wel00}. The distance to the source is
estimated to be anywhere between 4 kpc to 6.5 kpc \citep{Rut01}. 
The recurrent transient X-ray outbursts observed are thought to
occur due to thermal instabilities in the accretion disk \citep{Van96}.
In Aql X-1 the typical outburst recurrence time is
200 days, with average durations of 40-60 days \citep{Sim02}.
The source also shows Type-I X-ray bursts which are
believed to be run away thermonuclear burning of the matter accreted 
onto the surface of the NS \citep{Lew95,Bil98}. High frequency quasi periodic 
oscillations (QPOs) between 750-830 Hz and burst oscillations have 
also been discovered in this source \citep{Zha98}. Recently an episode 
of coherent pulsations at 550.27 Hz lasting for 150s was discovered 
during the peak of the 1998 outburst \citep{Cas08}. The occurrence of 
Type-I bursts, burst oscillations and the coherent pulsations in 
Aql X-1 confirm that the compact object is a rapidly spinning neutron star. 
\cite{Rei04} analysed five X-ray outbursts of Aql X-1
observed using RXTE-PCA between 1997 February to 2002 May. They
conclude from their timing and spectral studies that this is an Atoll
source.

Here we study the broad band spectrum of Aql X-1 using the pointed
\emph{\suzaku} and \emph{Rossi X-ray Timing Explorer} (RXTE) observations.

\begin{figure}
\begin{minipage}{80mm}
\centering
\includegraphics[height=80mm, angle=-90]{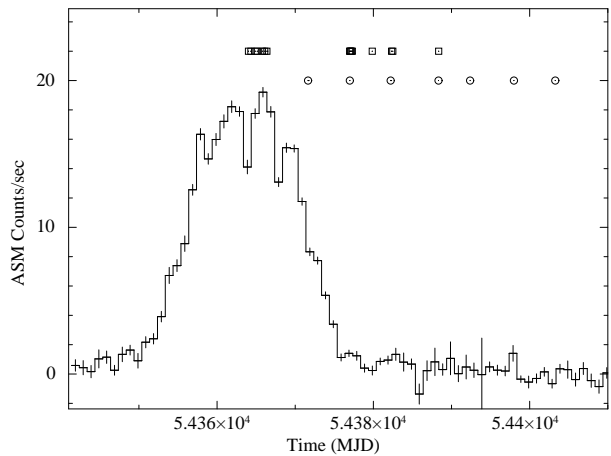}
\caption{Figure shows the outburst light curve of Aql X-1 as observed using 
the RXTE-ASM (1.5-12 keV). The square points marked at the top of the 
figure show the times of RXTE-PCA observations and the circled points below 
the squares show the times of \emph{Suzaku} observations.We note here
that a flux of 1 Crab corresponds to $\sim 75$ ASM counts/s \citep{Lev96}.}
\label{fig:asm-lc}
\end{minipage}
\end{figure}

\section{Observation and analysis}
Aql X-1 went into outburst during August 2007 and was observed using
\emph{Suzaku} \citep{Mit07} and RXTE \citep{Jah96}. The source flux 
increased from $\sim 10$ mCrab to $\sim 0.2$ Crab during the outburst
before it finally decays to a flux of $\sim 5$ mCrab. 
Figure \ref{fig:asm-lc} shows the one day averaged RXTE All Sky Monitor 
\citep[ASM;][]{Lev96} light curve of this outburst. Seven pointed 
observations are made using \emph{Suzaku}. The X-ray Imaging 
Spectrometer \citep[XIS;][]{Koy07} was operated in the 1/4 window mode
during all observations. An additional 0.5s burst option was used
during the first observation (Obs Id 402053010). Since all these 
observations were done after November 9, 2006 no XIS2 data
are available. Hard X-ray Detector \citep[HXD;][]{Tak07} observations
were done with XIS nominal pointing. For the last two \emph{Suzaku} observations
the count rates are low and the spectra obtained from XIS and HXD
have poor statistics. The fifth \emph{Suzaku} observation data
could not be analysed since the attitude file provided by the HEASARC 
site is corrupted. Hence these three observations are not used 
for the work presented here. For more details see Table \ref{tab:table1}. 

There are 16 pointed RXTE Proportional Counter Array (PCA) 
observations between 54360 MJD and 54400 MJD, seven during 
the peak of the outburst and nine after the decay of the outburst 
(Figure \ref{fig:asm-lc}). Three of the PCA observations are 
simultaneous to the second, third and fourth Suzaku observations 
(Table \ref{tab:table1}). In the following subsections we give the analysis 
details of data reduction for RXTE PCA and \emph{Suzaku} respectively. 

\subsection{RXTE-PCA}

\begin{figure}
\begin{minipage}{80mm}
\centering
\includegraphics[height=70mm, angle=-90]{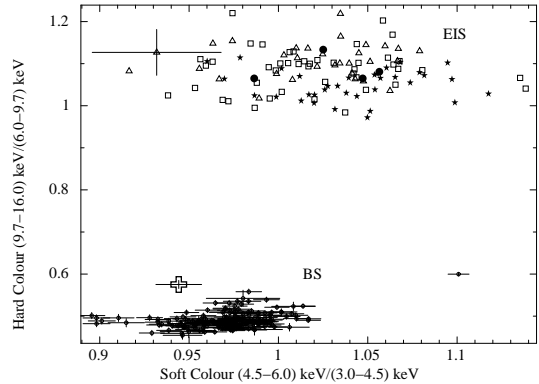}
\caption{Figure shows Colour-Colour diagram of Aql X-1 for the 2007 outburst. 
The open circles with error bars are points during the peak of the outburst 
where no \emph{Suzaku} observations are available. Points marked with filled 
circles, filled stars and triangles are from RXTE observations simultaneous 
to the second, third and fourth \emph{Suzaku} observations respectively. Note 
that there are no RXTE observations close to the first \emph{Suzaku} observation. 
Points marked with rectangles are remaining RXTE observations after the decay 
of the outburst where no simultaneous Suzaku observations are available. 
The point marked with an open plus sign and error bars is obtained from the PCA 
spectra simulated using the best fit model for the first \emph{Suzaku} 
observation. Error bars on points representing the EIS of Aql X-1 are 
suppressed for clarity except one point in the upper left hand corner.}
\label{fig:cd}
\end{minipage}
\end{figure}

PCA data are used to create the CD of Aql X-1 for the 2007 outburst.
To normalise the Aql X-1 colours by Crab colours we used
the three PCA Crab observations between 54360 MJD and 54400 MJD.
Background subtracted lightcurves of Aql X-1 and Crab were extracted using 
Std2 mode data in energy bands 3.0-4.5 keV, 4.5-6.0 keV, 6.0-9.7 keV, 
9.7-16.0 keV with 16s time resolution. To eliminate artifacts that may 
arise due to different gain corrections for the 
different PCUs only photons collected in the third PCU were used since this PCU 
was on during all the observations. The Soft colour (SC) is defined as the 
ratio of count rates in energy bands 4.5-6.0 keV and 3.0-4.5 keV and the
Hard colour (HC) as the ratio of count rates in energy bands 9.7-16.0 keV and 
6.0-9.7 keV. Average Crab colours obtained are 
SC = $1.986 \pm 0.002 $ and HC = $0.5992 \pm 0.0007$. Figure 
\ref{fig:cd} is the CD obtained by plotting normalised Aql X-1 HC versus 
SC. All normalised Aql X-1 colours having 
error greater than 5\% of the observed colour were ignored. 
This amounted in rejecting 14 out of 307 points of the CD. 
From Figure \ref{fig:cd} we conclude that 
during the second, third and fourth \emph{Suzaku} observations, Aql X-1 was
in the EIS 
\citep[compare Figure \ref{fig:cd} of this paper with Figure 2 of][]{Rei04}
We will refer to these three 
\emph{Suzaku} observations as EI1, EI2, EI3 respectively. 
It is noted that EI2 observation has the best statistics in the EIS and also
EI2 is brighter than EI1 indicating irregular decline in the 
lightcurve (Table \ref{tab:table1}).

There are no simultaneous PCA observations during the first 
\emph{Suzaku} observation to ascertain the spectral state of
the source. Hence we use a different method to find the normalised 
Aql X-1 SC and HC during this observation. The best fit 
spectral model that can reproduce the observed \emph{Suzaku} 
spectrum is used to simulate a PCA spectrum. The \texttt{fakeit}
command of XSPEC package is used for this simulation. 
Average count rates in the four energy bands required to calculate
the SC and HC are obtained by this simulated PCA spectrum. 
Aql X-1 SC and HC thus obtained are normalised by the 
average Crab SC and HC respectively. In figure \ref{fig:cd} the point 
marked by an open plus sign with error bars represents 
this normalised Aql X-1 SC and HC. Thus we conclude 
that during the first \emph{Suzaku} observation, Aql X-1 was in the
BS. Henceforth this observation will be referred to as the BS 
observation.

The PCA spectra from simultaneous observations (Table \ref{tab:table1})
are used for comparison with the \emph{Suzaku} spectra. We have used the
Std2 spectra and response files provided in the standard products. 
Energy channels were appropriately regrouped 
before the final spectral fitting.

\subsection{\emph{Suzaku}}

For each front-illuminated XIS detector (XIS0 and XIS3) we
extracted the source spectra using a 260 arc-sec circular 
extraction region centered on the source. The energy scale 
is reprocessed using the \texttt{xispi} task. Background
spectra are extracted using appropriate circular regions
outside the source region. Corresponding response files are
generated using the \texttt{xisrmfgen} and \texttt{xissimarfgen}
tools. The spectra from respective XIS0 and XIS3 detectors are
then added together using the \texttt{addascaspec} tool. 

\begin{figure}
\begin{minipage}{80mm}
\centering
\includegraphics[height=70mm, angle=-90]{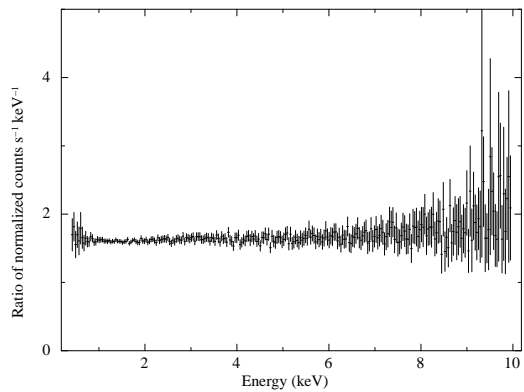}
\caption{Figure shows a plot of the ratio of XIS0 spectrum for
BS observation, obtained using a full circular source region of radius
260 arc-seconds to the spectrum obtained from an annular region of
inner radius 40 arc-seconds and outer radius of 260 arc-seconds at
different energies. The constant ratio implies that both these spectra
differ only in the observed count rates and there is no pile-up at the
centre of the CCD during the BS observation.}
\label{fig:compare-xis0}
\end{minipage}
\end{figure}

During high count rates, as is the case for the BS observation,
there is a possibility of pile-up in the XIS detectors. To check
for pile-up we extracted the source spectrum from an annular region
with inner radius of 40 arc-sec and an outer radius of 260 arc-sec.
Corresponding response files were also generated. Figure 
\ref{fig:compare-xis0} shows the ratio of XIS0 background 
subtracted spectrum extracted from the full circular region and
from the annular region. The constant ratio indicates
that both these spectra are similar and differ only in the
observed count rates. Similar result is seen for XIS3 detector also. 
This confirms that the BS observation is not affected by pile-up.

The data processing version used for the HXD cleaned products
is 2.1.6.15. Hence the PIN spectra are extracted using the
cleaned event files following the standard analysis threads 
on the \emph{Suzaku} website.\footnote[1]{http://heasarc.nasa.gov/docs/suzaku/analysis/abc/}
The PIN non-X-ray background 
is extracted from the observation specific model provided
by the instrument team. It is combined with the standard 
model for cosmic X-ray background. Appropriate response files
for XIS nominal pointing available from the calibration 
database CALDB (version XIS-20101108, HXD-20101202) are 
used.\footnote[2]{http://heasarc.gsfc.nasa.gov/docs/heasarc/caldb/suzaku/docs/\\suzaku$\_$caldbhistory.html}

The XIS spectra are grouped such that the final spectra have $\sim 250$
energy channels so that there are about three channels per energy resolution of $\sim 150 eV$. 
Energy channels corresponding to the energy range of
1.75-1.95 keV are ignored since the response files do not adequately
remove the absorption edge present in this energy range. 
Ten consecutive energy channels are grouped into one energy 
channel of the final grouped PIN spectrum. The net count rates 
for the XIS and PIN spectra are given in Table \ref{tab:table1}. 
We use \texttt{XSPEC} V12.5.1n to fit the spectral models 
to the observed X-ray spectrum.

\section{Results}

\begin{table}
\begin{minipage}{80mm}
\caption{XSPEC Model Components used}
~\\
\begin{tabular}{ll}
\hline
Model & XSPEC Components \\
\hline
M1 (a/b) & C$^{\dag}$ $\times$ wabs (diskbb + diskline + nthcomp)$^{a,b}$\\
M2 & C$^{\dag}$ $\times$ wabs (BBody/BBodyrad + nthcomp) \\
\hline
\end{tabular}
\label{tab:XSPEC_models}
~\\
\dag : Constant used to get relative normalisation between XIS and HXD spectra \\
a: The input photons to nthcomp are from accretion disk \\
b: The input photons to nthcomp are from an hidden blackbody emitting NS surface
\end{minipage}
\end{table}

\begin{table*}
\begin{minipage}{180mm}
\begin{center}
\caption{Parameter values with respective $1 \sigma$ errors for model M1a during the EIS}
~\\
\begin{tabular}{lllllll}
\hline
M1a Parameter                  & \multicolumn{2}{c}{EI1}              & \multicolumn{2}{c}{EI2} 
                               & \multicolumn{2}{c}{EI3}                       \\
	                       & 402053020             & 93405-01-04-01         & 402053030            & 93405-01-05-01    
                               & 402053040             & 93405-01-06-00        \\
\hline
$\rm{nH}(10^{22}) \rm{cm}^{-2}$&$0.311 \pm 0.008$      & $0.311 \rm{(frozen)}$  & $0.287 \pm 0.007$    & $0.287 \rm{(frozen)}$
                               &$0.445 \pm 0.009$      & $0.445 \rm{(frozen)}$ \\
$T_{in} \rm{(keV)}$            &$0.85  \pm 0.02 $      & $ 0.85 \pm 0.11 $      & $0.82  \pm 0.02 $    & $0.97  \pm 0.09     $
                               &$0.82  \pm 0.02 $      & $ 0.79 \pm 0.05 $     \\
$L_{DBB} \rm{(10^{35} erg/s)}$ &$5.1  \pm 2.4 $        & $ 5.0 \pm 2.5 $        & $4.7  \pm 2.2 $      & $6.9  \pm 3.2     $
                               &$5.2  \pm 2.5 $        & $ 6.9 \pm 3.3 $       \\
$R_{in}^{\dag} \rm{(km)}$      &$6.6  \pm 0.2 $        & $6.8  \pm 1.1 $        & $6.7  \pm 0.2 $      & $5.9  \pm 0.9     $ 
                               &$7.1  \pm 0.2 $        & $9.0  \pm 1.1 $       \\
$\gamma$                       &$1.71  \pm 0.04 $      & $1.85  \pm 0.05 $      & $1.78  \pm 0.03 $    & $1.73  \pm 0.04     $
                               &$1.71  \pm 0.05 $      & $1.74  \pm 0.04 $     \\
$\rm{kT_e (keV)}$              &$20.0  \rm{(frozen)}$  & $20.0  \rm{(frozen)}$  & $20.0  \rm{(frozen)}$& $20.0  \rm{(frozen)}$
                               &$13.6^{+4.9}_{-2.3}$   & $9.7^{+2.9}_{-1.7}$   \\ 
$L_{comp} \rm{(10^{35} erg/s)}$&$25.5 \pm 12.1$        & $25.4 \pm 12.1$        & $29.4 \pm 14.0$      & $30.6 \pm 14.5    $
                               &$21.8 \pm 10.4$        & $19.8 \pm 9.4$        \\
$L_{inp} \rm{(10^{35} erg/s)}$ &$4.6  \pm 2.2 $        & $6.1 \pm 2.9$          & $6.1  \pm 2.9$       & $6.1 \pm 2.9$
                               &$4.3  \pm 2.0$         & $4.6 \pm 2.2$         \\
$A$                            & $5.54$                & $4.16$                 & $4.82$               & $5.01$ 
                               & $5.07$                & $4.30$                 \\
$C\ddag$                       & $1.05 \pm 0.06$       & -                      & $1.11 \pm 0.05$      & -
                               & $1.10 \pm 0.07$       & -                      \\
$\chi^2/\nu$                   & $266.70/240$          & $30.48/28$             & $260.93/236$         &$37.03/28$
                               & $287.22/240$          & $24.23/23$             \\
\hline
\end{tabular}
\end{center}
$\dag$ Propagating distance errors in calculating error on $R_{in}$ would increase the errors to $\sim 5$ km for all observations.\\
$\ddag$ C is the cross-calibration constant between the XIS and HXD spectra.
\label{tab:EIS-M1a}
\end{minipage}
\end{table*}

We have used two different phenomenological models to describe the observed spectra of the source
in both the EIS and BS. 
 In the first model (henceforth referred to as M1) we model the soft component using multi-temperature 
disk blackbody (DBB) and the hard component as Comptonised
emission from the boundary layer. The hot boundary layer is assumed to completely cover the neutron star
surface.  This Eastern-like model has two variants. One, where
the input photons to the boundary layer are from the accretion disk (M1a)
and second where the input photons are from the underlying  blackbody emission of the neutron
star surface (M1b). The presence of a standard accretion disk allows for the possibility of
a broad Iron line emission in this model which we include using the \texttt{XSPEC} model ``diskline''.
In the second Western-like model (M2) we model the soft component using a single-temperature blackbody (BB)
emitted from the boundary layer. The hard component is modelled as
Comptonised emission from an inner hot disk where the input photons are from the
boundary layer.

The Comptonised spectra for both the models is represented by the \texttt{XSPEC} function ``nthcomp'' \citep{Zdz96, Zyc99}.
We apply this Comptonization model mainly because it can have input seed photon populations of different nature and 
distribution (pure blackbody, multicolour disk) thus allowing to test different emission geometries. In the M1a model 
the input photon distribution is due to disk blackbody and hence the input temperature is set equal to the
maximum temperature of the disk blackbody emission. In the M2b model, input photon distribution is due to 
the blackbody emission from the NS surface and hence the input temperature is a free parameter. On the other hand
in the M2 model, the photon distribution is due to neutron star boundary layer and hence of a blackbody shape and the input
tempertature is set equal to the temperature of this neutron star boundary layer.

A multiplicative component (``wabs'') is introduced in all spectral fitting to 
model the absorption of the source spectrum by
intervening material along the line sight. We note here that the average Dickey \& Lockman
nH value in the direction of Aql X-1 is $0.33 \times 10^{22}$ cm$^{-2}$.
The nH values obtained in our fits agree with this average value.
Another constant multiplicative term is used in both the models for the relative
normalisation of the XIS and corresponding HXD spectra. The cross-calibration constant
for XIS nominal pointing as suggested by the Suzaku team is $1.16 \pm 0.014$
\footnote[3]{http://www.astro.isas.jaxa.jp/suzaku/doc/suzakumemo/suzaku\\memo-2007-11.pdf}$^,$\footnote[4]{http://www.astro.isas.jaxa.jp/suzaku/doc/suzakumemo/suzaku\\memo-2008-06.pdf}. 
The value for the constant that we get from our fits agree with those suggested by the Suzaku team within errors. The
\texttt{XSPEC} model components used for M2 and M1 are tabulated in
Table \ref{tab:XSPEC_models}. While fitting the near simultaneous
{\it RXTE} spectra,   we fix
the absorbing column density, nH to the corresponding nH value
derived from the \emph{Suzaku} spectral fit since {\it RXTE} data alone
cannot constrain the nH value.

The normalisation parameter of the \texttt{XSPEC BBody} component
gives the blackbody luminosity $L_{BB}$ and the normalisation
parameter of the \texttt{XSPEC BBodyrad} component gives the blackbody
radius $R_{BB}$. Similarly from the parameter values of the
\texttt{XSPEC nthcomp} component we can estimate the total Compton
luminosity $L_{comp}$, the required luminosity of the input photons
$L_{inp}$, the amplification factor (A) and when $kT_{bb}$ is a free
parameter the radius of the blackbody $R_{BB}$. From the \texttt{XSPEC
diskbb} model we calculate the inner accretion disk radius ($R_{in}$)
and the disk luminosity $L_{DBB}$.  The calculated values of these
parameters allow us to constrain the two models considered here. Note
that throughout this paper for calculating luminosities we use distance
to the source as $D = 5.25 \pm 1.25 $ Kpc (Rutledge et
al. 2001) and the error quoted for luminosities take into account this
uncertainty in the distance.

The results from the observations EI1, EI2 and EI3 when the
source was in the EIS will be presented
together in the next sub-section and that of the BS 
in the subsequent one. 

\subsection{Spectral analysis of Aql X-1 during the Extreme Island State}

\begin{figure*}
\begin{minipage}{170mm}
\centering
\includegraphics[height=75mm, angle=-90]{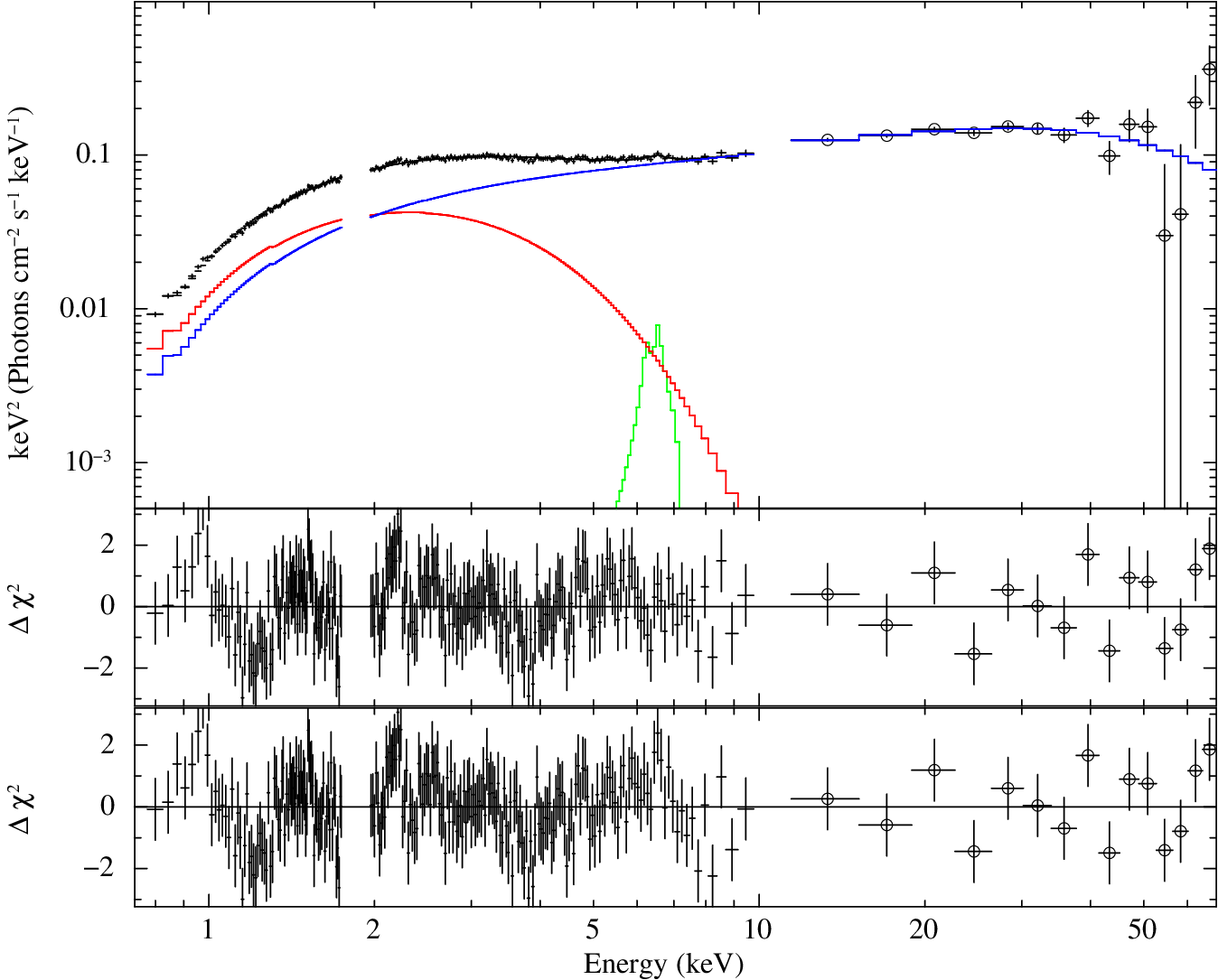}
\hspace{1cm}
\includegraphics[height=75mm, angle=-90]{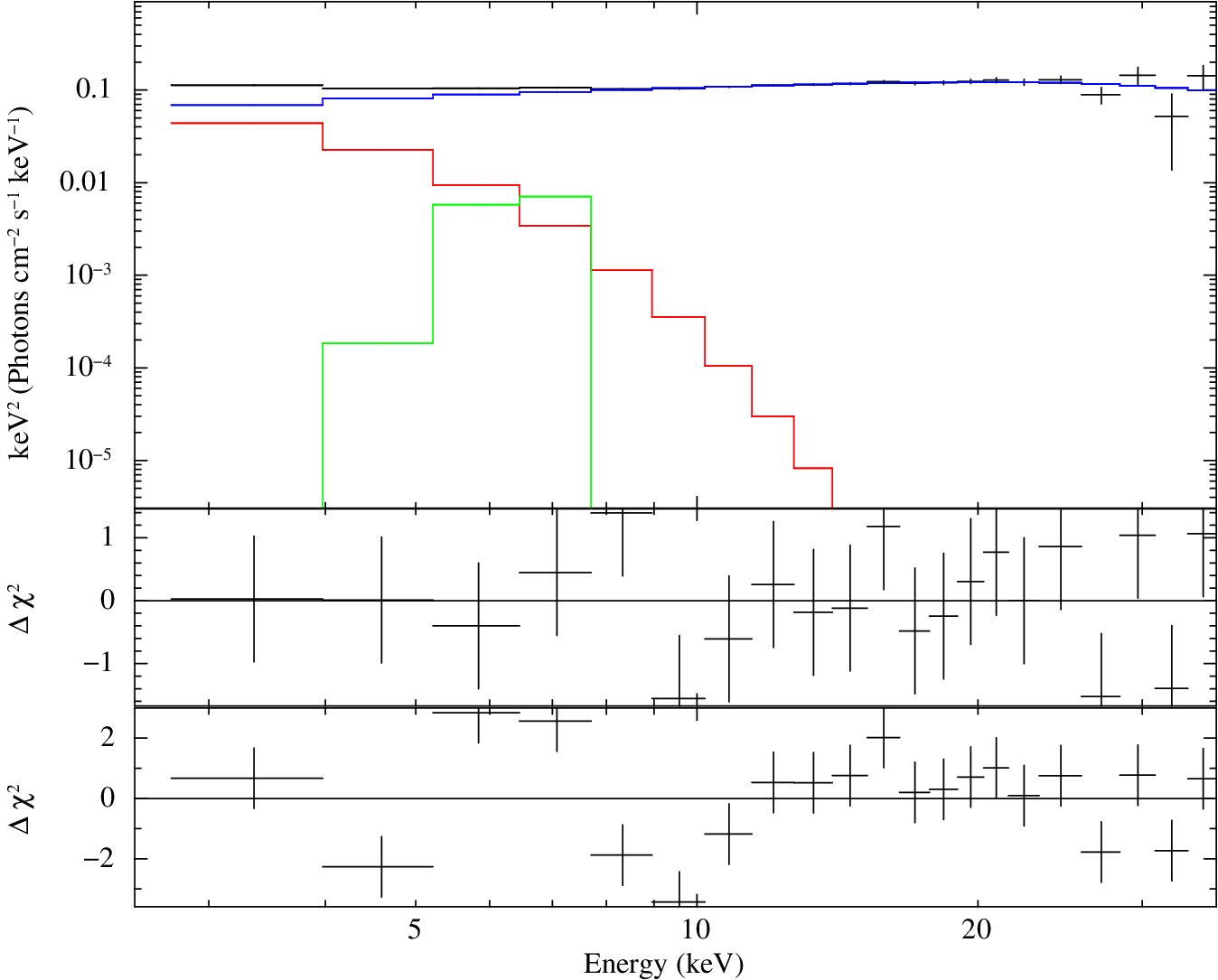}
\caption{The left hand panel shows the \emph{Suzaku} and the right
hand panel shows the corresponding RXTE-PCA unfolded spectrum for the
EI3 observation. Observed data points are marked in black. In the left
panel points below 10 keV are from the XIS detector and the points
marked with a circle are from the HXD detector. Red line shows the
contribution from the disk blackbody component, the Blue line shows
the contribution of the Comptonised component and the green line shows
the contribution from the Fe line. The bottom panels show the residues as
a function of energy with (middle panel) and without (bottom panel) the Fe line. }
\label{fig:M1a-ufspec}
\end{minipage}
\end{figure*}

We start with analysing the \suzaku spectra using Model M1a. If the
Iron line emission is omitted the best fit
$\chi^2/\nu$ obtained are 284.50/241, 295.74/237 and 305.31/241 for
observations EI1, EI2 and EI3 respectively. Adding a Gaussian line at $6.4$ keV improves
the fit considerably. The width of the Gaussian turns out to be $> 0.5$ keV, indicating the
line is broad. Hence, we adopt the more physical ``diskline'' model to represent the
Iron line. However, the data is not good enough to constrain all the parameters of the
model and hence we fix some of them (emissivity index $\beta = -2.0$, inclination angle
$i = 45^\circ$, inner radius $R_{in} = 10 \rm{ GM/c^2}$ and outer radius $R_{out} = 1000 \rm{ GM/c^2}$). 
Thus only the normalisation of the line is a fitted parameter. The
same model is applied to near simultaneous {\it RXTE} observations.  Removing the Iron
line component for the {\it RXTE} spectra, gives best fit $\chi^2/\nu$ values of 38.39/29, 70.08/29,
64.21/24 for Obs. Id. 93405-01-04-01/05-01/06-00 respectively, which are significantly higher than
 $\chi^2$ obtained with the line feature. Thus, the Iron line feature is detected both in the
\suzaku as well as in the {\it RXTE} spectra. 

The electron temperature of the Comptonizing boundary layer component, $kT_{in}$ is unconstrained for
observations EI1 and EI2 for both \suzaku and {\it RXTE} data with a lower limit of above 15keV.
Hence  we fix $kT_{in} = 20$ keV 
for these observations. The spectral parameters and the best
fit $\chi^2/\nu$ obtained for the \suzaku and {\it RXTE} spectra  are given in Table
\ref{tab:EIS-M1a}.  Figure
\ref{fig:M1a-ufspec} shows the $E^2F(E)$ spectrum for \emph{Suzaku}
observation EI3 and corresponding RXTE spectrum along with the M1a model components. Note that
the best fit spectral parameters are consistent within errors for the \suzaku and {\it RXTE}
observations for all three sets. 
This similarity and consistency between the two different instruments, shows that the spectral
extraction and response generation is robust and correct.

The inner radius of the accretion disk turns out to be $\sim 7$ km which is rather small since
it needs to be larger than the radius of the neutron star $R_s \sim 10$ km. However, this
may be due to the uncertainty in the distance to the source. Assuming that the distance 
is $6.5$  instead of $5.25$ kpc used here would lead to physically acceptable inner radii of $> 10$ km.  
Moreover, the colour factor assumed here is $f = 1.7$ and a slightly higher value could also correct
this discrepancy. Note that the luminosity of the disk is of the same order as the input luminosity
into the Comptonizing medium. This is an important consistency check since the
model assumes that input photons arise from the accretion disk.

Next we consider the possibility that the input photons to the Comptonizing
medium arise from a blackbody emission of the underlying neutron star surface
(Model M1b). Thus in this model temperature of the blackbody emission, $kT_{bb}$ is an extra free parameter. The best fit $\chi^2/\nu$ values are 268.35/239, 266.18/235 and 260.83/239 for EI1, EI2 and EI3 observations respectively. The {\it RXTE} spectral fits are also consistent with the \suzaku fits, although the extra freedom of $kT_{bb}$ does not allow the spectral parameters to be well constrained. But it is noted that for EI1 and EI2, the values of $kT_{bb}$ and $T_{in}$ coincide within errors whereas for EI3 $kT_{bb} < T_{in}$ which is not compatible with the geometry of this model. The calculated inner disk radii values for this model turn out to be larger at $\sim 9$ km. However, the radius of the blackbody assumed to be from the neutron star surface, turns out to be rather small ($\sim 5$ km) for the first  two observations, namely EI1 and EI2. Again, physical consistency may be obtained by using a larger distance to the source and/or including a colour factor in the blackbody radius calculation. Also for the EI3 observation, the obtained radius of the blackbody is $R_{BB} = 18 \pm 5$ km which is greater than the calculated inner disk radius $R_{in} = 7.5 \pm 0.3$ km. Therefore, perhaps the more physical picture for EIS is that the input photons for Comptonization are mainly contributed by the accretion disk making M1a model more feasible than the M1b model in this case.

We consider next the alternate geometry where the Comptonised component
arises from an inner hot disk and the boundary layer produces a blackbody
emission (Model M2). When this model is used to fit the three EIS spectra the best fit
$\chi^2/\nu$ for observations EI1, EI2 and EI3 are 300.87/241, 306.83/237
and 267.50/241 respectively. In
this model the input photon temperature for Comptonization component
$kT_{bb}$ is kept equal to the corresponding blackbody temperature
$kT$. The calculated blackbody radius $R_{BB}$ for the three
observations are $3.6 \pm 0.9$, $3.5 \pm 0.8$ and $4.1 \pm 1.0$
km respectively. These values are significantly smaller than the neutron star radius of 
$\sim 10$ km. The values can be made consistent only by assuming a large distance
to the source of $> 11$ kpc or by assuming an unphysical large colour factor of $\sim 3$.
Note that for a radius of 3 km the emitting region is only one-tenth the neutron star surface. 
This makes the geometry unfavourable as compared to the M1 model.

Moreover, the $\chi^2$ values obtained are significantly larger than the ones obtained
for M1 models. Even if we include a broad Iron line (which in the absence of a cold disk
may not be physical) the   $\chi^2/\nu$ values are 293.99/239, 285.49/235 and 260.46/239
which are higher than the ones obtained for the M1 model. Consistent results are obtained
from the {\it RXTE} data which provide $\chi^2/\nu$ of 34.40/28, 87.43/29 and 24.33/23
respectively, again significantly higher than what was obtained earlier.

\subsection{Spectral analysis of Aql X-1 during the Banana State}

We now come to the spectral analysis of the first
\emph{Suzaku} observation taken during the decline of the 2007 Aql X-1
outburst. During this observation the overall X-ray source flux is
higher than that during the next three observations when the source is
in the EIS. It was seen that for this BS \emph{Suzaku}
spectrum there is an excess of count rate at the high energy end of
the spectrum which could not be modelled by thermal Comptonization. 
We have added a hard power-law component (with photon index 
$\Gamma = 0$) to all spectral fitting to account for the high energy excess.
We note here that energy contribution of this component diverges with
energy and hence there should be a cutoff at both the high and low energy
end. But for the given data neither the photon index nor the cutoffs at
low and high energies could be constrained. Therefore luminosity contributed by this
component has been calculated only in the energy range of 0.7-70.0 keV 
in which the spectral fitting has been performed.

\begin{figure}
\begin{minipage}{80mm}
\centering
\includegraphics[height=73mm, angle=-90]{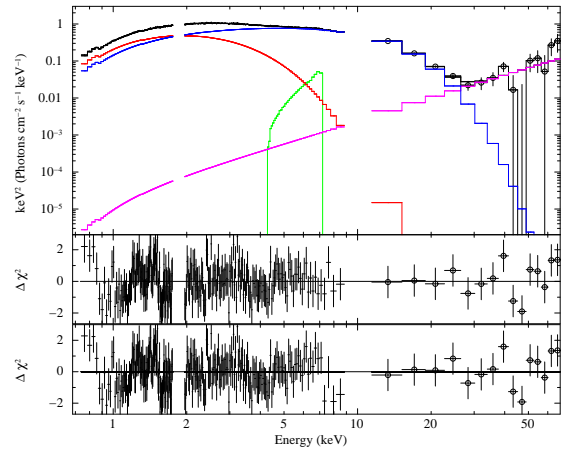}
\caption{The M1a model spectral components (listed in Table \ref{tab:obsid10-M1} along with the 
unfolded
spectrum for the BS \emph{Suzaku} observation. Black
points below 10 keV show the XIS spectrum and black points marked with
circles show the HXD spectrum. 
The multicoloured disk blackbody component is  shown in
red, the Comptonised component is shown in blue, the power-law
component in magenta and the Fe line component in green. The bottom panels
show residues as a function of energy with (middle panel) and without (bottom panel)
Fe line.}
\label{fig:HF-ufspec}
\end{minipage}
\end{figure}

Table \ref{tab:obsid10-M1} lists the best fit parameters for the variations
of the cold accretion disk model (M1a/b) while  Figure
\ref{fig:HF-ufspec} shows the unfolded spectrum with the different
components for the M1a model. The electron temperature of
the Comptonizing region is smaller ($kT_e \sim 2.3$ keV) as expected and  the inner radius of the
accretion disk is significantly larger at $\sim 40$ km. As in the
EIS fitting, the disk luminosity is nearly equal to the input luminosity (as
expected in this geometry), although the luminosities have increased by nearly
an order of magnitude. Here, in M1b model, the radius of the underlying neutron
star, $R_{BB} \sim 10$ km, which is larger (and hence more physical) than that
obtained for the EIS state. Note that the high energy power-law component 
contributes around $1.5$\% of the total luminosity. Without this component
the $\chi^2/\nu$ is 282.81/234 for M1a model and 279.90/233 for M2b model.
For both M1a and M1b model, the emissivity index of the ``diskline'' model can be 
constrained to $q \sim -3$.

The alternate geometry of an hot inner disk (M2) provides an unacceptably 
larger $\chi^2/\nu$ of 359.99/235. The inclusion of a broad Gaussian for the
Iron line (which may not be physical given the absence of a cold disk) reduces it to 
338.43/233, which is still significantly higher than what is obtained for
the M1 models. However, in this case the radius of the blackbody emitting
boundary layer turns out to be $\sim 18$ km and hence is physically possible.

\begin{table}
\caption{Parameter values of model M1(a/b) for Obs Id 402053010}
~\\
\begin{tabular}{lll}
\hline
M1 Parameter & M1a Values & M1b Values \\
\hline
nH($10^{22}$) cm$^{-2}$ & $0.370 \pm 0.009$               & $0.366 \pm 0.008$ \\
$T_{in}$(keV)           & $0.66 \pm 0.02$                 & $0.71 \pm 0.02$ \\
$L_{DBB}$ (erg/s)       & $(5.8 \pm 2.8) \times 10^{36}$  & $(10.3 \pm 4.9) \times 10^{36}$ \\
$R_{in}$ (Km)$^a$       & $37.1 \pm 1.3 $                 & $ 43.1 \pm 1.7 $ \\
$\Gamma^b$              & $0.0 \rm{(frozen)}$             & $0.0 \rm{(frozen)}$\\
$L_{PL}^c$ (erg/s)   & $(2.6 \pm 1.2) \times 10^{35}$  & $(2.7 \pm 1.3) \times 10^{35}$\\
$\gamma$                & $1.94 \pm 0.09 $                & $2.1_{-0.3}^{+0.6}$ \\
kT$_e$ (keV)            & $2.3 \pm 0.1$                   & $2.4^{+0.2}_{-0.1}$ \\ 
kT$_{bb}$ (keV)         & = T$_{in}$                      & $1.1^{+0.3}_{-0.4}$\\ 
$L_{comp}$ (erg/s)      & $(10.6 \pm 5.0) \times 10^{36}$ & $(6.1 \pm 2.9) \times 10^{36}$\\
$L_{inp}$ (erg/s)       & $(4.6 \pm 2.2) \times 10^{36}$  & $(4.3 \pm 2.1) \times 10^{36}$\\
A                       & 2.30                            & 1.41 \\
$R_{BB}$ (Km)$^d$       & -                               & $9.5 \pm 6.9$ \\
q                       & $-3.1^{+0.6}_{-1.7} $           & $-3.1^{+0.7}_{-4.7} $ \\
$ C $                   & $1.19 \pm 0.09$                 & $1.16 \pm 0.08$ \\
$\chi^2/\nu$            & 246.78/233                      & 246.10/232 \\ 
\hline
\end{tabular}
~\\
$a$: Propagating error on distance increases error in $R_{in}$ to 25 km for M1a and to 29 km for M1b.\\
$b$: The power law should have both a low and high energy cut-off which cannot be constrained by the present data.\\
$c$: Luminosity due to the powerlaw tail is calculated in the energy range 0.7-70.0 keV. See results section for more details. \\
$d$: Propagating error on distance increases error in $R_{BB}$ to 7.2 km. \\

\label{tab:obsid10-M1}
\end{table}

\section{Summary and Discussion}

In this work, we analyse \suzaku and {\it RXTE} 
observations of Aql X-1, to constrain the geometry of the source in both the low flux 
EIS and high flux BS, and to report the discovery of a hard X-ray tail in the BS.
 In particular, a standard accretion disk with a hot Comptonizing boundary layer whose
input photons are from the disk is the preferred geometry for both states.
The alternate model of a hot Comptonizing inner disk and a blackbody emitting boundary
layer gives significantly larger $\chi^2$ for all four \suzaku observations including
three nearly simultaneous {\it RXTE} observations. Moreover, for the EIS 
observations, the radius of the blackbody emitting boundary layer turns out to be
unphysically small, $\sim 3$ km. The presence of a broad Iron line feature 
provides further support for the existence of a cold accretion disk in the source.
The preferred geometry is the same as inferred from broad band study of many 
other NS-LMXBs \citep{Bar01} and hence seems to be general.

For all observations, the luminosity of the input photons to the 
Comptonizing hot boundary layer is similar to that of the disk and hence it
is possible that the main source of input photons is the disk itself. 
 While it may also be possible that the source of soft photons could 
be the underlying blackbody emitting neutron star, statistically the data cannot
differentiate between these scenarios.

Our results confirm that the electron temperature of the Comptonizing boundary layer 
in EIS is higher as compared to same in the BS. Also the inner radius of the accretion
disk is higher during the BS observation. The spectral slope $\gamma \sim 1.8$ 
remains nearly constant, implying that the optical depth has increased in the high 
flux state. In the ``nthcomp'' model, the optical depth of the medium, for a spherical geometry can be estimated 
by \citep{Sun80}
\begin{equation}
  \Gamma_{thcomp} = \left[\frac{9}{4} + \frac{1}{(kT_e / m_e c^2) \tau (1 + \tau / 3)}\right]^{1/2} - \frac{1}{2}
  \label{eq_tau}
\end{equation} 
The optical depth $\tau \sim 2$ for the observations EI1 and EI2 and $\sim 5$ for the EI3
observation. In the BS $\tau \sim 13$.  The difference in the
electron temperature is the primary cause for the two states to have different locations in
the colour-colour diagrams. A larger number of observations, sampling the source at different
times of the outburst will provide more information regarding the complete temporal evolution
of the source.

For the EISs, when the input photons are assumed to be
from the disk (Table \ref{tab:EIS-M1a}), the luminosity of the disk
is $\sim 5 \times 10^{35}$ ergs s$^{-1}$. The luminosity of the boundary
layer, which is the luminosity in the Comptonised component  minus the
input luminosity $L_{comp}-L_{inp}$, is $\sim 2.5 \times 10^{36}$ ergs s$^{-1}$. Thus
energy released in the boundary layer is about $5$ times larger than
that of the disk.  When the
input photons are taken to be the underlying neutron star surface, boundary layer luminosity, 
$L_{comp} \sim 2.8 \times 10^{36}$ ergs s$^{-1}$ is about a factor of four more than the disk luminosity
$\sim 7 \times 10^{35}$ ergs s$^{-1}$. For the BS this ratio is $\sim 1$ when the
input photons are disk photons and $\sim 0.7$ when they arise from the underlying star surface.  
Theoretically, the ratio of the energy output in the
boundary layer to that of the disk can range from $1$-$3$ depending on the
spin and equation of state of the neutron star \citep[e.g.][]{Bha00}. A ratio of
five would imply an extremely soft equation of state. However, the simple phenomenological
models used here do not warrant one to make concrete statements but one can
conclude that the boundary layer is more luminous than the disk, which is consistent
with the expected energy release ratio.

An interesting result is the detection of an hard X-ray power-law emission in the high flux
state which is only $1.5$\% of the total source luminsoity. Such hard tails have been
detected in Z sources and only one Atoll source GX 13+1 earlier \citep{Pai06}. 
The origin of this emission is unknown and maybe due to some non-thermal emission in a jet 
\citep{Mig06}. Indeed, a weak radio emission has been detected in the high soft state of 
Aql X-1 \citep{Rup04}, which is also observed in other atoll sources \citep{Mig06}. 
Alternatively, the hard X-ray emission could be due to an hybrid non-thermal emission \citep[e.g.][]{Cop99} or
bulk motion Comptonization \citep[e.g.][]{Tit97}.
Multiple observations of the source may reveal whether the hard X-ray tail  is variable 
and whether it is correlated with the radio emission.

The work highlights the utility of broad band spectral analysis of LMXB. Apart from
other \suzaku observations the upcoming Indian satellite ASTROSAT will be useful in such studies
of LMXBs containing weakly magnetised neutron stars as it will provide good timing and
spectral resolution.

\section{Acknowledgements}

{}

\end{document}